\documentclass[aps,groupedaddress,twocolumn,showpacs,showkeys,superscriptaddress]{revtex4-1}
\usepackage{bm,amsmath}
\usepackage{graphicx}
\usepackage[font={small}]{caption}
\usepackage[caption=false]{subfig}
\usepackage{float}
\usepackage{epstopdf}

\begin{document}

\title{Pattern formation in reaction-diffusion systems in the presence of short-term memory}

\author{Reza Torabi}
\email{rezatorabi@aut.ac.ir} 
\affiliation{Department of Physics and Astronomy, University of Calgary, Calgary, Alberta, T2N 1N4, Canada}
\affiliation{Department of Physics, Tafresh University, 39518-79611, Tafresh, Iran}

\author{J\"orn Davidsen}
\email{davidsen@phas.ucalgary.ca}
\affiliation{Department of Physics and Astronomy, University of Calgary, Calgary, Alberta, T2N 1N4, Canada}
\affiliation{Hotchkiss Brain Institute, University of Calgary, Calgary, Alberta T2N 4N1, Canada}

\begin{abstract}
 
We study reaction-diffusion systems beyond the Markovian approximation to take into account the effect of memory on the formation of spatio-temporal patterns. Using a non-Markovian Brusselator model as a paradigmatic example, we show how to use reductive perturbation to investigate the formation and stability of patterns. Focusing in detail on the Hopf instability and short-term memory, we derive the corresponding complex Ginzburg-Landau equation that governs the amplitude of the critical mode and we establish the explicit dependence of its parameters on the memory properties. Numerical solution of this memory dependent complex Ginzburg-Landau equation as well as direct numerical simulation of the non-Markovian Brusselator model illustrate that memory changes the properties of the spatio-temporal patterns. Our results indicate that going beyond the Markovian approximation might be necessary to study the formation of spatio-temporal patterns even in systems with short-term memory. At the same time, our work opens up a new window into the control of these patterns using memory.

\end{abstract}

\maketitle

\section{Introduction}

The term ``non-Markovian process'' covers all stochastic processes with the exception of the small minority that happens to have the Markov property \cite{van Kampen}. Despite the fact that stochastic processes in nature are generally non-Markovian, they are usually treated in the context of a Markov approximation. In this approximation, it is assumed that the correlation time between the system and environment is infinitely short so that memory effects can be neglected. Actually, treating non-Markovian processes with a Markov approximation can be as poor as, for instance, approximating highly nonlinear dynamical systems by a harmonic oscillator \cite{van Kampen}. Since realistic systems typically possess a finite correlation or scattering time, they are non-Markovian by nature and considering memory effects is often inevitable.

Specifically, non-Markovian processes appear in many different fields including quantum optics \cite{Breuer,Gardiner,Lambropoulos}, solid state physics \cite{Lai}, quantum chemistry \cite{Shao,Pomyalov}, quantum information processing \cite{Aharonov,Maniscalco}, and even in the description of biological systems \cite{Rebentrost}. Strongly coupled systems are also non-Markovian where the collisions can not be considered  as instantaneous \cite{Prigogine2}. Recently, attention has been paid to memory effects in non-Markovian systems in different areas \cite{Schumann,Trimper,Serva,Fritzsche,Harris,Dalmaroni}. For instance, taking into account long-range memory in extreme events by using fractional L\'evy processes \cite{Watkins, Moloney} allows one to make a time-dependent hazard assessment of future events based on events observed in the past \cite{Schumann}. The effect of memory-dependent transport on the survivability of a population is investigated in \cite{Harris}. The extension of conventional diffusive transport, within the framework of evolution equations, is considered in \cite{Trimper} to take into account memory. Memory effects in correlated anisotropic diffusion are studied in nanoporous crystalline solids in \cite{Fritzsche}. In this paper we take into account memory in reaction-diffusion systems to investigate its effect on the formation of patterns out of equilibrium.

Reaction-diffusion systems are extensively used in the study of self-organized phenomena that occur in open systems out of equilibrium \cite{Nicolis,Cross,Desai,Kerner,Buceta,Mimura,Dividsen}. They are useful in many fields such as biology \cite{Harrison,Meinhardt,Murray1,Murray2,Kariga,Chaplain,Sherratt1,Sherratt2,Gatenby,Kondo,Shin}, chemistry \cite{Kapral,Dividsen2,Beta,Castets,Ruan,Lengyel,Gafiychuk,Horvath,Horvath2,Porjai,Showalter,Malchow}, medicine \cite{Cameron,Fenton,Echebarria,Alonso}, neuroscience \cite{Liang,Lefevre,Tuckwell,Habib,Wu}, physics \cite{Astrov,Arecchi,Staliunas,Nakoa} and ecology \cite{Holmes,Skellam,Gowda,Meron}. Reaction-diffusion systems were first proposed by Alan Turing in the study of morphogenesis \cite{Turing}. Actually, Turing noticed that adding a diffusion term to a reaction system can drive the system to instability and plays an important role in the formation of patterns out of equilibrium. The characteristic feature of most of the studied reaction-diffusion systems is that the diffusion is considered to be a normal one. From microscopic point of view, the normal diffusion is derived from the (Markovian) master equation \cite{Reichl,Srokowski}.

The master equation is one of the most important equations in statistical physics that governs the dynamics of stochastic processes. As a system of stochastic variables evolves in time, transitions occur between various states of the system. To formulate the master equation, it is assumed that the probability of each transition depends only on the preceding time step and not on any previous history, which is exactly the Markov approximation. Since ``normal'' transport theory employs the master equation as its point of departure, normal diffusion does not take into account memory effects. The significance of the diffusion term in the formation of patterns and the importance of understanding and controlling pattern formation in far from equilibrium systems --- an area of significant importance \cite{Bertram,Mikhailov,Zhang2,Cui,Hu,Hata,Li,Zhang3,Hu} --- motivates us to study pattern formation in reaction-diffusion systems in the presence of memory. Using analytical and numerical analysis, we show that memory can change the properties of spatio-temporal patterns. This includes the stability of spiral waves, which occur in many chemical and biological systems such as the heart \cite{Bar,Fenton2,FentonH,Bernus,Dividsen}. This opens up a new mechanism to control pattern formation using memory instead of external perturbations such as applying periodic forcing \cite{Zhang2,Cui} or time-delayed feedback \cite{Hu,Li,Zhang3}.

The paper is organized as follows. In section II we construct non-Markovian reaction-diffusion systems by starting from the generalized master equation to take into account memory effects. Linear stability analysis is used in section III to investigate the effect of memory on the instability in a non-Markovian Brusselator model as a paradigmatic example of non-Markovian reaction-diffusion systems. In section IV, reductive perturbation method is exploited to study the behavior of the system near a Hopf instability and investigate the effect of memory on spatio-temporal patterns. The results indicate that a complex Ginzburg-Landau equation that depends on the memory governs the amplitude equation of the critical mode for short-term memories. Section VI contains the numerical solution of the obtained complex Ginzburg-Landau equation as well as direct numerical simulations of the non-Markovian Brusselator model to illustrate the effect of memory on the formation of spatio-temporal patterns. Finally, a summary and discussion is presented in section VI.

\section{Non-Markovian reaction-diffusion systems}

The normal diffusion equation, ${\partial n(x,t)}/{\partial t}=D {\nabla^{2} n(x,t)}$, is a macroscopic equation where $ n(x,t)$ is the density of the substrate at point $x$ and time $t$. This equation can be obtained starting from the master equation \cite{Reichl}. In this section we are going to establish a general equation governing non-Markovian reaction-diffusion systems. For this purpose, we first review how to introduce a non-Markovian diffusion equation by starting from the generalized master equation. 

Various methods have been used to obtain the generalized master equation, however, the procedure of Zwanzig \cite{Zwanzig,Zwanzig2} distinguishes itself by its elegance and economy of effort. His procedure was based on projection operator techniques \cite{Nakajima,Zwanzig,Zwanzig2,Prigogine}, which are widely used in open systems and non-equilibrium statistical mechanics \cite{Haake,Balescu,Grabert,Kubo}. Zwanzig started from the Liouville-von Neumann equation for the density operator, and obtained the generalized master equation \cite{Zwanzig, Zwanzig2,Kenkre,Srokowski} 
\begin{equation} \label{GME}
\dfrac{\partial P_\xi(t)}{\partial t}= \int_0^t dt' \sum_{\mu} [W_{\xi\mu}(t-t')P_{\mu}(t')-W_{\mu \xi}(t-t')P_{\xi}(t')],
\end{equation}
where $P_\xi$  are the diagonal elements of density matrix and denote the probability of finding the system in state $\xi$. $W_{\xi \mu}$ are the transition rates from state $\mu$ to $\xi$ that generally are time dependent. Integration over all previous times indicates the non-Markovian nature of the processes. Since this equation is hard to solve, physicists use Markov approximation to simplify this equation. By this approximation, they ignore the dependence of transition rates to previous times \cite{Kenkre} and consider
\begin{equation}
W_{\mu\xi} (t')=F_{\mu\xi} \delta(t'),
\end{equation}
where $F_{\mu\xi}$ are the transition rates in Markovian case which are time independent. By neglecting the dependence on previous times they assume that there is no memory in the system. As a result, the equation \eqref{GME} reduces to the well known master equation,
\begin{equation} \label{mastermarkov}
\dfrac{\partial P_\xi(t)}{\partial t}= \sum_{\mu} [F_{\xi\mu}P_{\mu}(t)-F_{\mu \xi}P_{\xi}(t)].
\end{equation}
Applying the nearest-neighbor approximation on the transition rates and taking the continuum limit \cite{Reichl,Kenkre,Srokowski}, one can derive 
\begin{equation} \label{diff}
\dfrac{\partial P(x,t)}{\partial t}=D\dfrac{\partial^{2} P(x,t)}{\partial x^{2}},
\end{equation}
where $P(x,t)$ is the probability of finding a particle in $x$ at time $t$ and $D$ is the diffusion coefficient. Generalizing this equation to three dimensions and noting that the density of the substrate is proportional to this probability, one can obtain the normal diffusion equation.

Now, if we want to be more precise and consider realistic phenomena beyond the Markovian assumption, we should start from the generalized master equation \eqref{GME}. However, it is hard to work with this equation. A simplification may often be possible whereby the time dependence of $W_{\xi\mu}$ is assumed independent of the states $\xi$ and $\mu$ \cite{Kenkre}:
\begin{equation} \label{Simp}
W_{\mu\xi} (t)=F_{\mu\xi} {\cal K}(t).
\end{equation}
Using this assumption, \eqref{GME} can be rewritten in a simpler form
\begin{equation} \label{mastermarkovsimp}
\dfrac{\partial P_\xi(t)}{\partial t}=\int_0^t dt' {\cal K}(t-t')\sum_{\mu} [F_{\xi\mu}P_{\mu}(t')-F_{\mu \xi}P_{\xi}(t')].
\end{equation}
The nature of transport is thus decided by the memory ${\cal K}$ as well as by the rates $F_{\xi\mu}$. By applying the nearest-neighbor approximation on the transition rates $F_{\mu \xi}$ and taking the continuum limit, one obtains the non-Markovian Fokker-Planck equation \cite{Kenkre}
\begin{equation} \label{diffmarkov}
\dfrac{\partial P(x,t)}{\partial t}=\int_0^t dt' {\cal K}(t-t') D \dfrac{\partial^{2} P(x,t')}{\partial x^{2}},
\end{equation}
where ${\cal K}(t')$ is the memory kernel. This equation expresses the fact that an interaction process is in general a non instantaneous event and thus nonlocal in time. The memory kernel describes the finiteness of the correlation or scattering time \cite{Manne,Kenkre2}. Obviously, for the Markovian case, the memory kernel is $ {\cal K}(t')=\delta(t')$ and  \eqref{diffmarkov} converts to \eqref{diff}. In the study of realistic systems one encounters different forms of the memory kernel. For instance, the excitation energy transfer between molecules indicates the presence of an exponential memory kernel \cite{Kenkre}. However, the description of reactions in polymer systems needs more complicated kernels \cite{Cherayil,Bandyopadhyay}.

Using \eqref{diffmarkov}, we can write a general two component non-Markovian reaction-diffusion system as
\begin{equation} \label{activ-inhabit}
\left\{
\begin{array}{cc}
\frac{\partial U({\bf r},t)}{\partial t}=f_1(U,V;\mu)+\int_0^t dt' {\cal K}_U (t-t') D_{U} \nabla^2 U({\bf r},t')\\
\frac{\partial V({\bf r},t)}{\partial t}=f_2(U,V;\mu)+\int_0^t dt' {\cal K}_V (t-t') D_{V} \nabla^2 V({\bf r},t')
\end{array}
\right.,
\end{equation}
which is a non-Markovian activator-inhibitor system. Here, $f_1$ and $f_2$ are reaction terms. $U$ and $V$ are the densities of activator and inhibitor, respectively. $D_U$ and $D_V$ are their corresponding diffusion coefficients, and ${\cal K}_U$ and ${\cal K}_V$ are the memory kernels associated with them. $\mu$ is the bifurcation parameter and, as $\mu$ varies, the system might move from a steady state to an oscillating or a patterned state via a Hopf or Turing instability, respectively\cite{Kuramoto}.

\section{Linear stability analysis}

In the previous section we constructed a general two component non-Markovian reaction-diffusion system. In this section we use linear stability analysis to see how the memory affects the critical behavior of the system. In order to study the problem analytically in more detail, we consider the Brusselator model with memory in the diffusion of the inhibitor. The Brusselator model is a two-component activator-inhibitor system and it is one of the paradigmatic models for nonlinear chemical systems \cite{Reichl}. This system in the presence of memory, in the diffusion of the inhibitor, takes the form
\begin{equation} \label{nMBrus}
\left\{
\begin{array}{cc}
\frac{\partial U({\bf r},t)}{\partial t} =& A-(B+1)U+U^2V+ D_{U} \nabla^2 U({\bf r},t) \\
\frac{\partial V({\bf r},t)}{\partial t} =& BU-U^2V+\int_0^t dt' {\cal K}_V (t-t') D_{V} \nabla^2 V({\bf r},t')
\end{array}
\right..
\end{equation}
Here, $U$ and $V$ are chemical concentrations that can vary in space and time. $A$ and $B$ are constants and B is considered as the control parameter of the system to generate patterns. The steady state solution of the non-Markovian Brusselator model, \eqref{nMBrus}, is $(U_0,V_0)=(A,B/A)$. As the system \eqref{nMBrus} is difficult to treat analytically in the presence of general memory and also many systems exhibit short-term memory, we focus on this latter case in order to present an analytical study of non-Markovian reaction-diffusion systems. To this end, we use a change of variable $z=t-t'$ to rewrite the integral in \eqref{nMBrus} as 
\[
\int_0^t dz {\cal K}_{V} (z) D_{V} \nabla^2 V({\bf r},t-z).
\]
In the presence of short-term memory only small $z$'s are important because the memory kernel, ${\cal K}_V (z)$ is almost zero for large $z$'s. So, we can expand $V({\bf r},t-z)$ up to the first order in $z$, $V({\bf r},t-z)=V({\bf r},t)-z\frac{\partial V({\bf r},t)}{\partial t}$, and rewrite \eqref{nMBrus} as

\begin{equation} \label{nMBrusShort}
\left\{
\begin{array}{cc}
\frac{\partial U({\bf r},t)}{\partial t}=& A-(B+1)U+U^2V+D_{U} \nabla^2U({\bf r},t)\\
\frac{\partial V({\bf r},t)}{\partial t}=& BU-U^2V+D_{V} \nabla^2\big{(}\int_0^t {\cal K}_V(z) V({\bf r},t) dz\\
&-\int_0^t z {\cal K}_V(z) \frac{\partial V({\bf r},t)}{\partial t} dz\big{)}
\end{array}
\right..
\end{equation}
To proceed further, we need to consider a specific form for the memory kernel. A widely used form for the memory kernel function is an exponential one \cite{Srokowski,Kenkre}
\begin{equation}\label{expkernel}
{\cal K}(t')=\gamma e^{-\gamma t'},
\end{equation}
where $\gamma$ measures the reciprocal of the characteristic scattering time or in other words the memory kernel decay time $(\tau=1/\gamma)$. Note that, the kernel becomes the delta function in the limit $\gamma\rightarrow\infty$ (the Markovian case). By considering this memory kernel, \eqref{nMBrusShort} takes the form
\begin{equation} \label{nMBrusShortSimp}
\left\{
\begin{array}{cc}
\frac{\partial U({\bf r},t)}{\partial t}=& A-(B+1)U+U^2V+D_{U} \nabla^2 U({\bf r},t)\\
\frac{\partial V({\bf r},t)}{\partial t}=& BU-U^2V+D'_{V}(t) \nabla^2 V({\bf r},t) \\ 
&-D''_{V}(t) \nabla^2 \frac{\partial V({\bf r},t)}{\partial t}
\end{array}
\right..
\end{equation}
Here, $D'_V(t)=D_V(1-e^{-t/\tau_V})$, $D''_{V}(t)=D_{V}(\tau_V-(t+\tau_V)e^{-t/\tau_V})$ and $\tau_V=1/\gamma_V$ is the characteristic memory kernel decay time for inhibitor. $D'_V(t)$ and $D''_V(t)$ are monotonically increasing functions of time with $0<D'_V<D_V$ and $0<D''_V<D_V \tau_V$. In the second relation in \eqref{nMBrusShortSimp}, $\frac{\partial V({\bf r},t)}{\partial t}$ has appeared on both sides. We replace the right one by the full expression for $\frac{\partial V({\bf r},t)}{\partial t}$. Since  $0<D''_V<D_V \tau_V$ and $\tau_V$ is small, if $D_V$ is not so large, we can truncate the process to obtain

\begin{equation} \label{firstloop}
\left\{
\begin{array}{cc}
\frac{\partial U}{\partial t}=& A-(B+1)U+U^2V+D_{U} \nabla^2 U \\
\frac{\partial V}{\partial t}=& BU-U^2V+D'_V(t) \nabla^2 V -B D''_V(t) \nabla^2 U \\
&+D''_V (t) \nabla^2 (U^2V) -D'_V(t) D''_V(t)  \nabla^2 (\nabla^2 V)
\end{array}
\right..
\end{equation}
We are interested in the long time behavior of the system to study the formation of patterns. On the other hand, in the case of short-term memory where $\tau_V$ is small, the exponential term in $D'_V(t)$ and $D''_V(t)$ goes to zero rapidly. Therefore, equation \eqref{firstloop} can be approximated by

\begin{equation} \label{longtime}
\left\{
\begin{array}{cc}
\frac{\partial U}{\partial t}=& A-(B+1)U+U^2V+D_{U} \nabla^2 U \\
\frac{\partial V}{\partial t}=& BU-U^2V+D_{V} \nabla^2 V -B D_{V}\tau_V \nabla^2 U \\
&+D_{V} \tau_V \nabla^2 (U^2V-D_V \nabla^2 V)
\end{array}
\right..
\end{equation}

In order to perform a linear stability analysis, small perturbations, $u({\bf r},t)$ and $v({\bf r},t)$, about the steady state $(U_0,V_0)$ are considered
\[
U({\bf r},t)=U_0+u({\bf r},t),\;\;\;\;\;\; V({\bf r},t)=V_0+v({\bf r},t).
\]
Inserting the above in Eq.~\eqref{longtime} and choosing the linear part, we obtain a linear equation governing the dynamics of perturbations. Using the normal mode ansatz 
\[\left( {{\begin{array}{cc} u \\ v \\ \end{array}}} \right)=\left( {{\begin{array}{cc} c_1 \\ c_2 \\ \end{array}}} \right) exp(\lambda t+i{\bf k}\cdot {\bf r}),
\]
in the linearized equations leads to  
\begin{equation} \label{growthpertubShort}
\left( {{\begin{array}{cc}
		a_{11} & a_{12} \\
		a_{21} & a_{22} \\
		\end{array}}}\right)
\left( {{\begin{array}{cc}
		c_1  \\
		c_2 \\
		\end{array}}} \right)={\bf 0},
\end{equation}
where, 
\begin{align*} \nonumber
a_{11} &= B-1-D_U k^2 -\lambda,\\
a_{12} &= A^2,\\
a_{21} &= -B-D_V \tau_V B k^2,\\
a_{22} &= -A^2-D_V k^2 -D_V \tau_V A^2 k^2-D_V^2 \tau_V k^4-\lambda.
\end{align*}
$\lambda$ is the perturbation growth rate, ${\bf k}$ is the wave vector and $k= |\bf{k}|$. The characteristic equation in the presence of memory takes the form
\begin{equation} \label{ChMemory}
\lambda^2+g(k,\tau_V) \lambda+ h(k,\tau_V)=0,
\end{equation}
where
\begin{multline} \nonumber
g(k,\tau_V)=1-B+A^2+
\big{[}D_U+D_V+D_V \tau_V A^2\big{]} k^2+D_V^2 \tau_V k^4,
\end{multline}
and
\begin{multline} \nonumber
h(k,\tau_V)=A^2 B(1+D_V \tau_V k^2)+\\
(1-B+D_U k^2)(A^2+D_V k^2+D_V \tau_V A^2 k^2+D_V^2 \tau_V k^4).
\end{multline}
Using the characteristic equation, one can find the perturbation growth rate, $\lambda$, in terms of $k$. Wave vectors that result in a nonnegative real part of the growth rate $\lambda$ are critical. Note that this equation depends on the memory properties of the system and in the limit of no memory ($\tau_V\rightarrow 0$) we recover the characteristic equation for the Markovian Brusselator model \cite{Torabi}. 

The steady state is linearly stable if and only if both $g(k,\tau_V)$ and $h(k,\tau_V)$ are non-negative for all $k$. Clearly, this stability condition can be violated in either of the following two ways:

1) $h(k,\tau_V)$ vanishes for some $k$, but $g(k,\tau_V)$ and $h(k,\tau_V)$ remain positive for all $k$. This condition together with ${\partial h(k,\tau_V)}/{\partial k}|_{k=k_{cT}}=0$ determine the critical parameter, $B_{cT}$, and also $k_{cT}$ for a Turing bifurcation.    

2) $g(k,\tau_V)$ vanishes for some $k$, but $g(k,\tau_V)$ and $h(k,\tau_V)$ remain positive for all $k$. This condition together with ${\partial g(k,\tau_V)}/{\partial k}|_{k=k_{cH}}=0$ leads to the critical values $k_{cH}$ and $B_{cH}$ for a Hopf bifurcation. The frequency of oscillation is $\omega=\pm \sqrt{h(B_{cH},k_{cH})}$.

Focusing on the latter, we find that ${\partial g(k,\tau_V)}/{\partial k}=0$ leads to a solution of $k=0$ as well as
\begin{equation} \label{k2}
k^2=-\dfrac{D_U+D_V+D_V \tau_V A^2}{2D_V^2 \tau_V}.
\end{equation}
Since the right hand of \eqref{k2} is always negative, it is not an acceptable solution and we obtain that
\begin{equation} 
k_{cH}=0, \;\;\;\;\;\;\;\ B_{cH}=1+A^2,
\end{equation} 
which means that the critical values for Hopf instability do not change when memory is present in the diffusion of inhibitor. On the other hand, since the characteristic function depends on the memory property of the system, the growth rates of the modes depend on memory as well. This affects the modulation of unstable modes near the Hopf bifurcation and, thus, influences the formation of spatio-temporal patterns. In fact, linear analysis only provides us with insight about the critical behavior of the system. However, in order to investigate the effect of memory on pattern formation, we need to go beyond the linear analysis. In the next section, we use a reductive perturbation method to investigate the effect of memory on the formation of spatio-temporal pattern near a Hopf instability. 

\section{Reductive perturbation method}

Near a bifurcation the neighboring modes of the critical mode, which have decay times of the same order of magnitude as the critical mode, play an important role in the long-time behavior of the system that results in the generation of patterns. According to the slaving principle \cite{Kuramoto}, near a bifurcation we are left with a couple of relevant dynamical variables whose time scales are distinguishably slower than the other dynamical variables so that the latter can be eliminated adiabatically using rescaled space-time coordinates. In this section, we investigate the non-Markovian Brusselator model analytically using the reductive perturbation method. The results indicate that memory changes the properties of the spatio-temporal patterns.

Starting from equation \eqref{longtime}, considering small perturbations about the steady state and keeping nonlinear terms as well as the linear terms results in the equation governing the fluctuations as
\begin{equation} \label{Nonlinearfluc}
\left( {{\begin{array}{cc}
		\dot{u} \\
		\dot{v} \\
		\end{array}}} \right)=
{\bf L}
\left( {{\begin{array}{cc}
		u \\
		v \\
		\end{array}}} \right)+{\bf N}+{\bf D},
\end{equation}
where
\begin{equation} \label{Jacobian}
{\bf L}=\left( {{\begin{array}{cc}
		B-1 & A^2 \\
		-B & -A^2 \\
		\end{array}}}\right)
\end{equation}
is the Jacobian matrix, ${\bf N}=\left( {{\begin{array}{cc}
		\beta  \\
		-\beta \\
		\end{array}}} \right)$ is the nonlinear part with $\beta=\frac{B}{A}u^2+2Auv+u^2v$ and ${\bf D}=\left( {{\begin{array}{cc}
		D_1  \\
		D_2 \\
		\end{array}}} \right)$ is the diffusion term with
\begin{align*} \nonumber	 
D_1=& D_{U}\nabla^2 u,\\   
D_2=& D_{V} (1+\tau_V A^2) \nabla^2 v +D_{V} \tau_V \big{(} B \nabla^2 u+ \\
&2 A \nabla^2 (uv)+\frac{B}{A} \nabla^2 u^2 + \nabla^2 (u^2 v)-D_V \nabla^2 (\nabla^2 v)\big{)}.
\end{align*}

We introduce a new bifurcation parameter by considering $\mu=\frac{B-B_{cH}}{B_{cH}}$. By this definition the steady state is stable for $\mu<0$ and is unstable for $\mu>0$. Near the instability, ${\bf L}$, ${\bf N}$, ${\bf D}$, $\left( {{\begin{array}{cc}
		u \\
		v \\
		\end{array}}} \right)$
and the eigenvalues of the Jacobian matrix can be expanded in powers of $\mu$ as
\begin{eqnarray}\label{expand}
&{\bf L}={\bf L_0}+\mu {\bf L_1}+\mu^2 {\bf L_2}+...,\nonumber \\
&{\bf N}={\bf N_0}+\mu {\bf N_1}+\mu^2 {\bf N_2}+... ,\nonumber \\
&\left( {{\begin{array}{cc}
		u \\
		v \\
		\end{array}}} \right)=\mu^{\frac{1}{2}}\left( {{\begin{array}{cc}
		u_1 \\
		v_1 \\
		\end{array}}} \right)+\mu \left( {{\begin{array}{cc}
		u_2 \\
		v_2 \\
		\end{array}}} \right)+\mu^{\frac{3}{2}} \left( {{\begin{array}{cc}
		u_3 \\
		v_3 \\
		\end{array}}} \right)+... ,\nonumber \\
&\lambda=\lambda_0+\mu \lambda_1+\mu^2 \lambda_2+...,\nonumber \\
\end{eqnarray}
where $\lambda_0= i\omega_0=iA={\bf U_L} {\bf L_0} {\bf U_R}$ and $\lambda_i=\sigma_i+
i\omega_i={\bf U_L} {\bf L_1} {\bf U_R}$. The vectors ${\bf U_R}$ and ${\bf U_L}$ are right and left eigenvectors of ${\bf L_0}$, respectively. ${\bf L_0}$ and ${\bf L_1}$ are given by
\begin{equation} \label{L0}
{\bf L_0}=\left( {{\begin{array}{cc}
		A^2 & A^2 \\
		-1-A^2 & -A^2 \\
		\end{array}}}\right), \;\;\;\; {\bf L_1}=(1+A^2)\left( {{\begin{array}{cc}
		1 & 0 \\
		-1 & 0 \\
		\end{array}}}\right),
\end{equation}
and the right and left eigenvectors of ${\bf L_0}$ are
\[
{\bf U_R}=\left( {{\begin{array}{cc}
		1 \\
		-1+iA^{-1} \\
		\end{array}}} \right), \;\;\;\; {\bf U_L}=\dfrac{1}{2} \bigg{(}1-iA,\;\;\;\; -iA\bigg{)}.
\]

We must get inside the neighborhood of the critical mode to consider just the important modes based on the slaving principle. We do this by rescaling the time and space variables as \cite{Kuramoto,Torabi}
\[
T=|\mu| t,\;\;\; R=|\mu|^{\frac{1}{2}}r=\epsilon r,
\]
where $\varepsilon^2 \chi\equiv\mu$ and $\chi=\rm sgn(\mu)$. Since we are interested in the supercritical Hopf bifuracation, $\chi$ is positive. Reductive perturbation method is a multi-scale method, so, we consider $\left( {{\begin{array}{cc}
		u \\
		v \\
		\end{array}}} \right)$ as a function of $t$, $T$, $R$. This means that we are dealing with the long-time, long-wavelength modes in their natural variables $T$ and $R$, and reserving $t$ for the overall
periodic motion (limit cycle). Substitution of \eqref{expand} into \eqref{Nonlinearfluc} and equating coefficients of different powers of $\varepsilon$, yields to a set of equations in the form of
\begin{equation}\label{homo}
\bigg{(}\dfrac{\partial}{\partial t}-{\bf L_0}\bigg{)}\left( {{\begin{array}{cc}
		u_\nu \\
		v_\nu \\
		\end{array}}} \right)={\bf B_{\boldsymbol\nu}}, \;\;\; \nu=1,2,...,
\end{equation}
where the first three $\bf B$'s are
\begin{align*}
&{\bf B_1}={\bf 0},\nonumber\\
&{\bf B_2}=\left( {{\begin{array}{cc}
		\frac{1+A^2}{A} u_1^2+2Au_1v_1 \\
		-\frac{1+A^2}{A} u_1^2-2Au_1v_1 \\
		\end{array}}} \right),\nonumber\\
&{\bf B_3}= -\bigg{(}\dfrac{\partial}{\partial T}-L_1+{\bf \hat{D}}\nabla^2_R\bigg{)} \left( {{\begin{array}{cc}
		u_1 \\
		v_1 \\
		\end{array}}} \right)+ \\
	&\left( {{\begin{array}{cc}
		\frac{2(1+A^2)}{A} u_1u_2+2A(u_1v_2+u_2v_1)+u_1^2 v_1 \\
		-\frac{2(1+A^2)}{A} u_1u_2-2A(u_1v_2+u_2v_1)-u_1^2 v_1 \\
		\end{array}}} \right),\nonumber
\end{align*}	
and ${\bf \hat{D}}$ is given by
\[
{\bf \hat{D}}= \left( {{\begin{array}{cc}
		D_U & 0 \\
		D_V \tau_V (1+A^2) & D_V(1+\tau_V A^2) \\
		\end{array}}}\right).
\]
Note that, in the limit of $\tau_V \rightarrow 0$ this matrix reduces to $\left( {{\begin{array}{cc}
		D_U & 0 \\
		0 & D_V \\ 
		\end{array}}}\right)$, which is identical to ${\bf \hat{D}}$ in Markovian Brusselator model \cite{Torabi}.
The solution for $\nu=1$ is
\[
\left( {{\begin{array}{cc}
		u_1 \\
		v_1 \\
		\end{array}}} \right)=W(T,R){\bf U_R}e^{i\omega_0 t}+c.c,
\]
where $c.c.$ stands for the complex conjugate and $W(T,R)$ is a complex amplitude to be determined. Using this solution in the equation for $\nu=2$ leads to an expression for $\left( {{\begin{array}{cc}
		u_2 \\
		v_2 \\
		\end{array}}} \right)$. Putting these two solutions into the equation for $\nu=3$ together with the solvability condition for the set of equations \eqref{homo} \cite{Kuramoto}, results in the equation governing the
	amplitude $W$
\begin{equation}\label{CoGLE}
\frac{\partial W}{\partial T}= \lambda_1 W+d\nabla^2_{\bf R} W-g|W|^2W,
\end{equation}
where $\lambda^{(1)}$, $d$ and $g$ are generally complex numbers and are given by

\begin{equation} \label{lambda1}
\lambda_1={\bf U_L} {\bf L_1} {\bf U_R}=\frac{1+A^2}{2},
\end{equation}
\begin{multline} \nonumber
d= {\bf U_L}{\bf D'}(\tau_U,\tau_V) {\bf U_R}=\\
\frac{1}{2}[D_U+D_V-A^2 D_V\tau_V-iA(D_U-D_V+D_V\tau_V)],
\end{multline}
\begin{equation} \nonumber
g=\frac{1}{2}(\frac{2+A^2}{A^2}+i\frac{4-7A^2+4A^4}{3A^3}).
\end{equation}
Equation \eqref{CoGLE} is the well known complex Ginzburg-Landau equation, but note that the coefficient $d$ depends on memory kernel decay time. In fact, the presence of non-Markovian diffusion of inhibitor results in the change in real and imaginary part of the coefficient $d$. For $\tau_V \rightarrow 0$, we recover the complex Ginzburg-Landau equation in Markovian Brusselator model \cite{Kuramoto,Torabi}. With a redefinition as follows
\begin{equation} \label{Rescaled}
r'= ({\sigma_1/d_r})^{1/2} R,\;\;\; t'=\sigma_1 T, \;\;\;
W'= \sqrt{g_r/\sigma_1}e^{-i\omega_1 T} W,
\end{equation}
the complex Ginzburg-Landau equation above the criticality can be written in a more convenient form (after dropping the primes)
\begin{equation}\label{CGLERS}
\frac{\partial W}{\partial t}= W+(1+ic_1)\nabla^2
W-(1+ic_2)|W|^2W,
\end{equation}
where 
\begin{equation} \label{c1}
c_1=d_i/d_r=-A\dfrac{D_U-D_V+D_V\tau_V}{D_U+D_V-A^2 D_V\tau_V},
\end{equation}
and 
\begin{equation} \label{c2}
c_2=g_i/g_r=\frac{4-7A^2+4A^4}{3A(2+A^2)}.
\end{equation}

Since the parameter $c_1$ depends on the memory property of the system (see \eqref{c1}), one can change the stability of solutions of the complex-Ginzburg Landau equation by changing the memory property of the system and, thus, control the resulting spatio-temporal pattern. In fact, by looking at the phase diagram of the complex Ginzburg-Landau equation \cite{Aranson}, one can see how the solutions of the complex Ginzburg-Landau and their stability change over the parameter space. By introducing memory or changing its properties (here by varying $\tau_V$), we can cross the boundaries of different regimes in any spatial spatial dimension, which enables us to change the selected spatio-temporal patterns. To explicitly illustrate how to achieve this and also to provide further support for our analytical results, we exploit numerical simulations in the next section using $\tau_V$ as our control parameter.

\section{Numerical simulations}

In this section, we perform numerical simulations of the complex Ginzburg-Landau equation obtained analytically in section IV as well as direct numerical simulations of the non-Markovian Brusselator model \eqref{nMBrus}, with the exponential memory kernel \eqref{expkernel}.  For integrating the complex Ginzburg-Landau equation we use a pseudospectral method to perform numerical computations in Fourier space based on the method of exponential time differencing (ETD2) \cite{Cox}. Small-amplitude random initial data about $W = 0$ and periodic boundary conditions are used. For the direct numerical simulation of the non-Markovian Brusselator model, we also consider small amplitude random initial condition about the steady state and use periodic boundary condition.  We chose to focus on two spatial dimensions, which displays a rich spectrum of dynamical behavior including spiral waves.

As a specific example, we focus on the stability of spiral waves. Fig. 1a shows rotating spiral waves that are generated in the absence of memory for the Markovian approximation of the Brusselator model, which is achieved by considering a delta function for the memory kernel in \eqref{nMBrus}. For the selected parameters and no memory, equations \eqref{c1}-\eqref{c2} give rise to $(c_1,c_2)=(-0.33,0.96)$ for the coefficients of the corresponding complex Ginzburg-Landau equation~\eqref{CGLERS}. As expected based on our analytical calculations, for these parameters the complex Ginzburg-Landau equation gives basically identical results as can be seen in Fig. 1c. Note that the change in scales between Fig. 1a and Fig. 1c is a consequence of rescaling \eqref{Rescaled} in obtaining \eqref{CGLERS}. Most importantly, $(c_1,c_2)=(-0.33,0.96)$ belong to the spiral bound states in parameter space \cite{Aranson}, consistent with the observed behavior. 
On the other hand, if we introduce memory in the Brusselator model, direct numerical simulations of \eqref{nMBrus} indicate that the selected pattern changes and spiral waves break up (see Fig.1b). In fact, by introducing memory with $\tau_V=1/\gamma_V=1/3.5$, the coefficient $c_1$  of the complex Ginzburg-Landau equation changes and the new coefficients $(c_1,c_2)=(-0.97,0.96)$ belongs to the defect turbulence regime beyond the absolute instability line where spiral waves are unstable \cite{Aranson,AransonA}. This is directly confirmed by Fig. 1d providing further support for our analytic calculations above. Thus, $\tau_V$ can be used as a control parameter to select stable or turbulent spiral patterns. 

\begin{figure}
	\begin{center}
		
		\subfloat[] {\includegraphics[width=0.49\columnwidth]{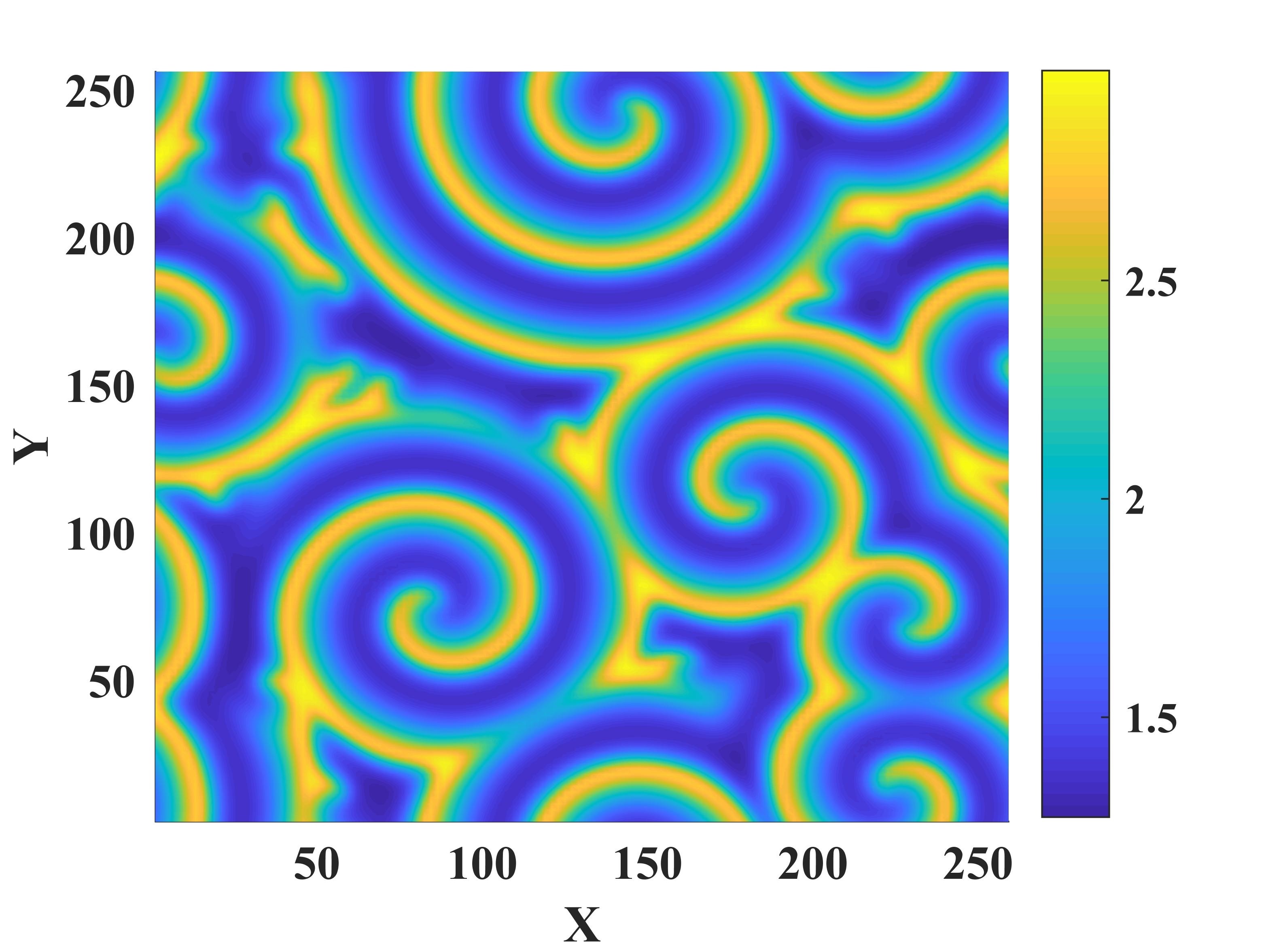}}
		\subfloat[] {\includegraphics[width=0.49\columnwidth]{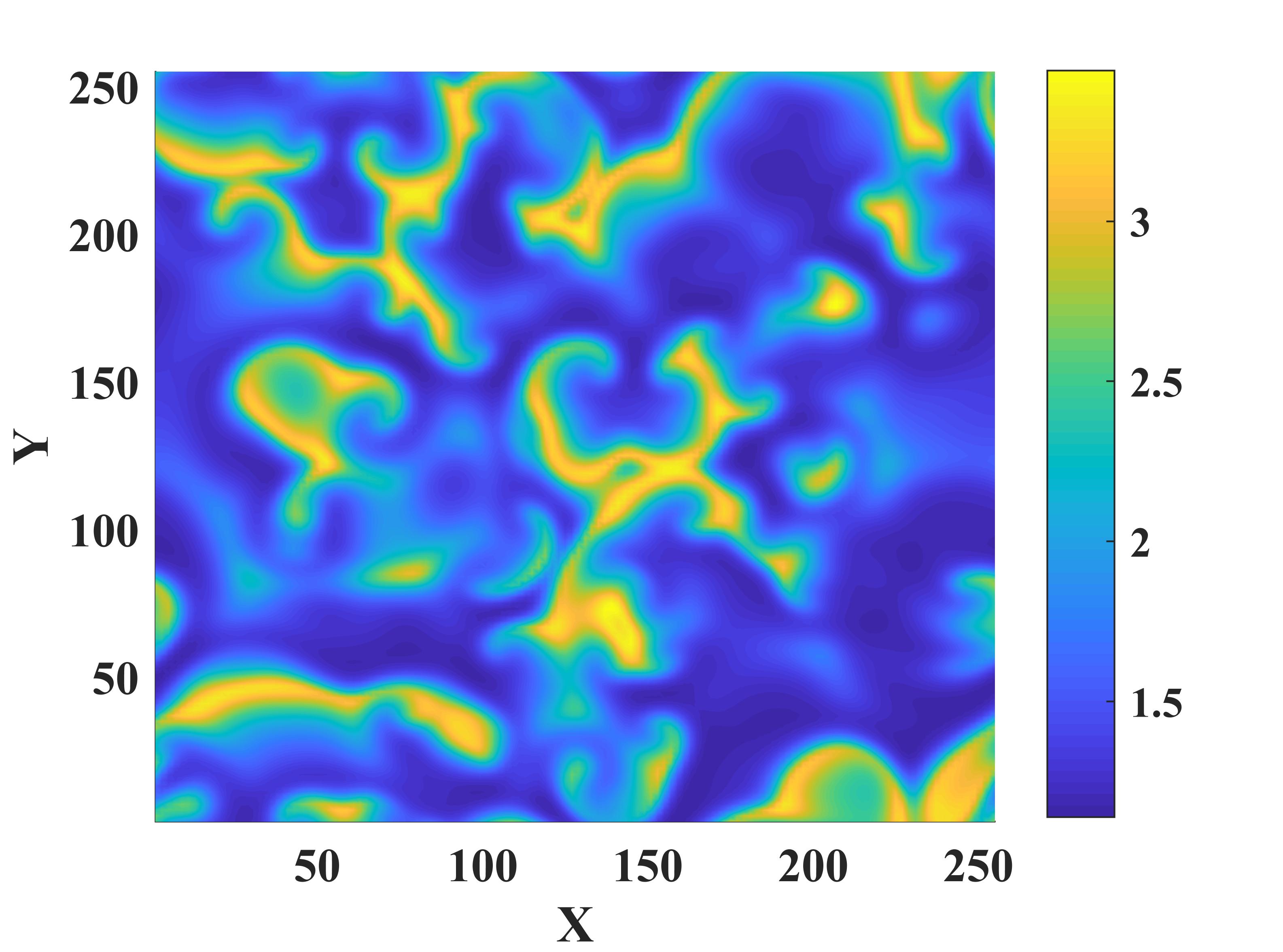}}\\
		\subfloat[] {\includegraphics[width=0.49\columnwidth]{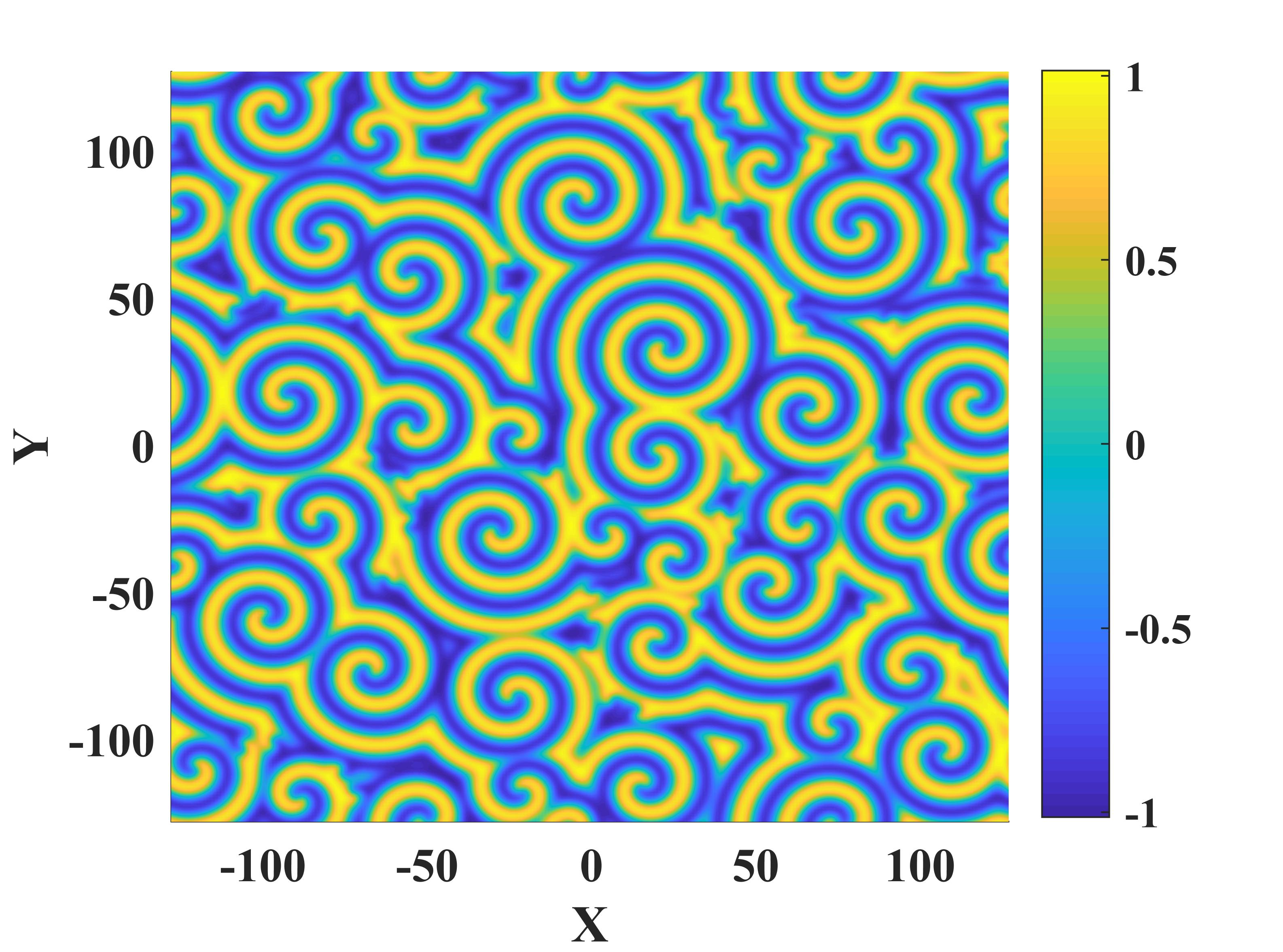}}
		\subfloat[] {\includegraphics[width=0.49\columnwidth]{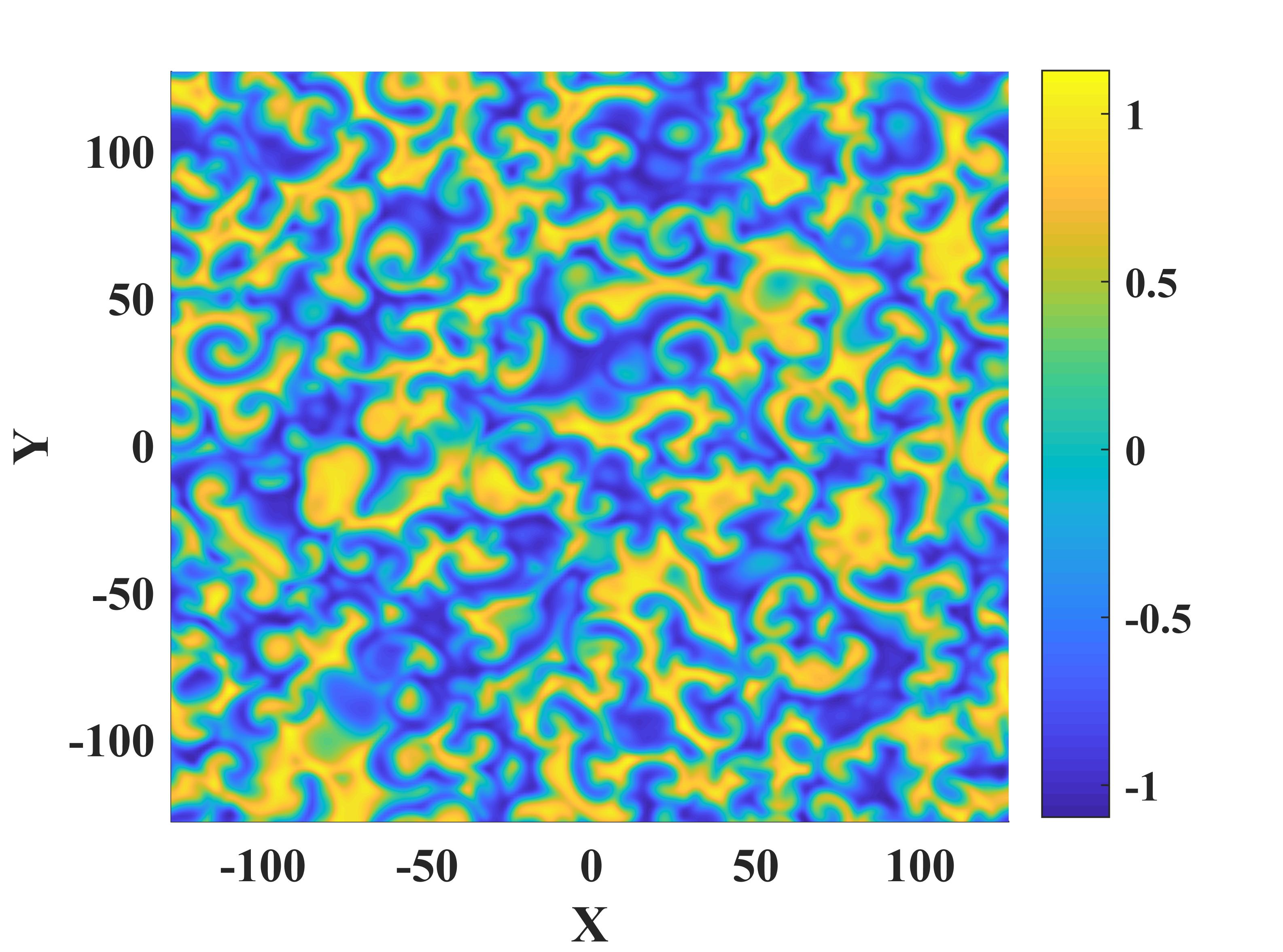}}\\
		\subfloat[] {\includegraphics[width=0.49\columnwidth]{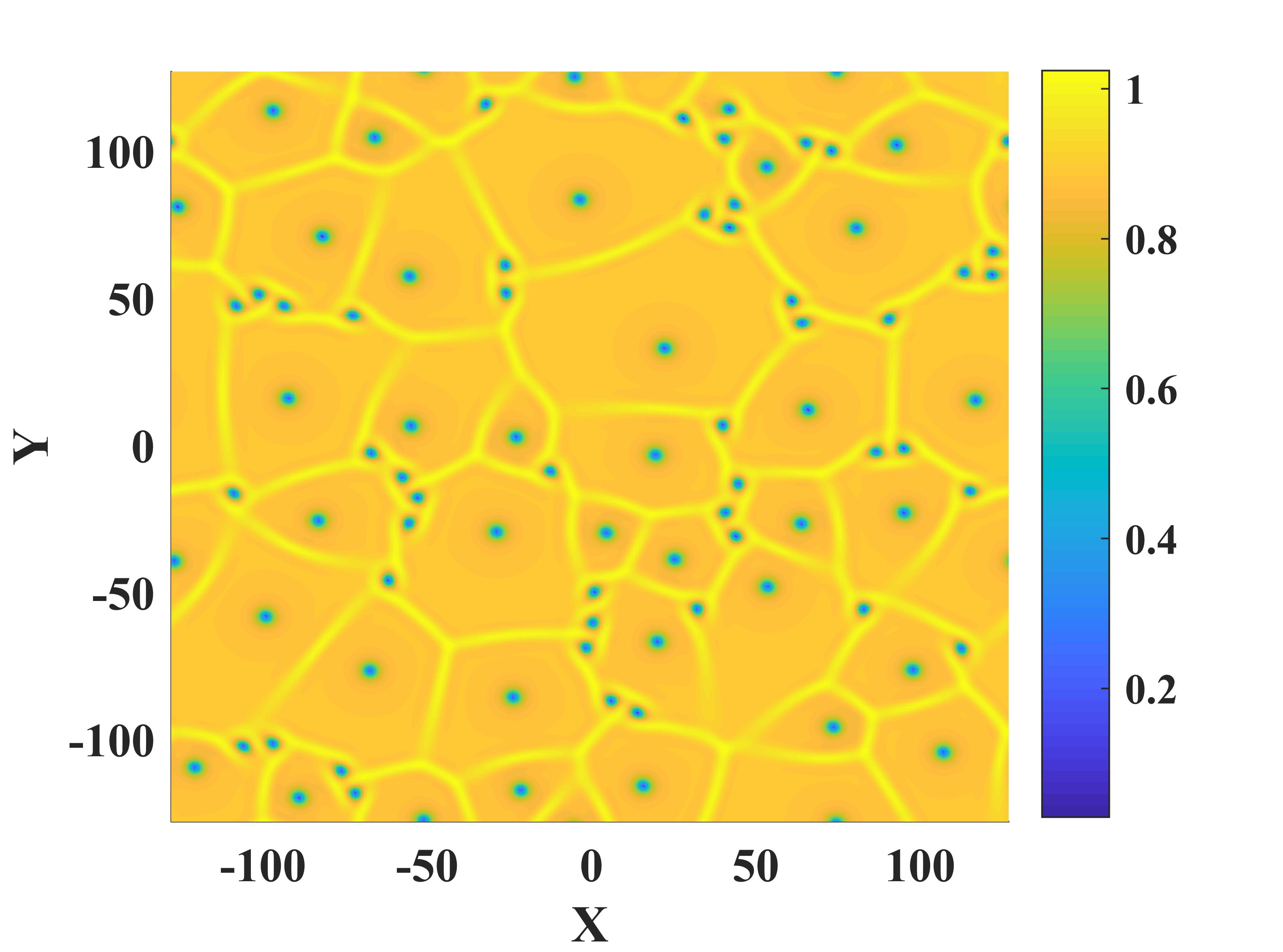}}
		\subfloat[] {\includegraphics[width=0.49\columnwidth]{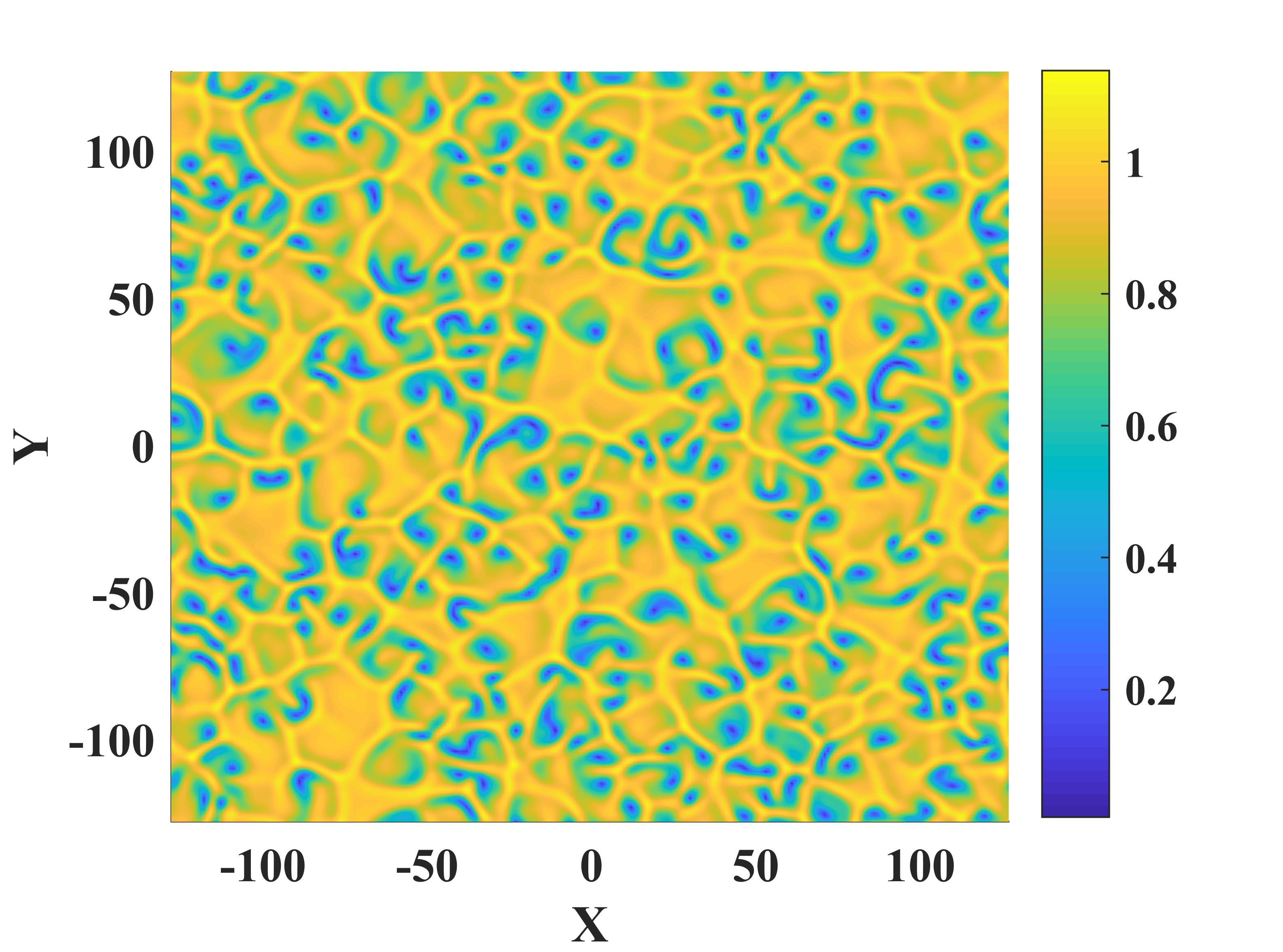}}\\
	\end{center}
        \caption{a) Snapshot of $U(x,y,t)$ in the Brusselator model~\eqref{nMBrus} in the absence of memory for parameters $A=1.9$, $B=4.8$, $D_U=1$, $D_V=0.7$ and a system size of $256\times256$. Rotating spiral waves appear in the system. b) Same as in a) but in the presence of memory with $\tau_V=1/\gamma_V=1/3.5$. Spiral waves break up when memory is turned on. c)-f) illustrative snapshots of the complex Ginzburg-Landau equation~\eqref{CGLERS} for parameters equivalent to those used in a) and b) using \eqref{c1} and \eqref{c2}. c) and e) show $Re(W)$ and $|W|$, respectively, in the absence of memory $(c_1,c_2)=(-0.33,0.96)$. d) and f) show $Re(W)$ and $|W|$, respectively, in the presence of memory $(c_1,c_2)=(-0.97,0.96)$.} \label{Spiralbreakup}
\end{figure} 

Changing the memory property of the system ($\tau_V$) and consequently changing the coefficient $c_1$ in the complex Ginzburg-Landau equation, not only helps us to break up spiral waves but it also enables us to change the properties of spiral patterns as well. To this end, one can start from the monotonic range \cite{Aranson,AransonK} and cross the oscillatory range line \cite{Aranson,AransonK} by changing $c_1$ to end up in the bound state regime. Fig.2 illustrates the behavior of the complex Ginzburg-Landau equation, obtained in \eqref{CGLERS}, for two sets of coefficients $(c_1,c_2)$ with different $c_1$. Fig. 2a shows $Re(W)$ for $(c_1, c_2)=(0.2,0.82)$ in the monotonic range where pairs of spirals form that drift and annihilate by interacting with each other. However, Fig. 2b corresponds to $(c_1,c_2)=(-0.33,0.82)$ in the spiral bound state regime of the oscillatory range. In order to illustrate the difference between the properties of the solutions, $|W|$ is shown in Fig. 2c and Fig. 2d. There are no shock lines in Fig. 2c for the monotonic range in contrast to Fig. 4d for the oscillatory range where cellular structures are formed. 

\begin{figure} 
	\begin{center}
		\subfloat[] {\includegraphics[width=0.49\columnwidth]{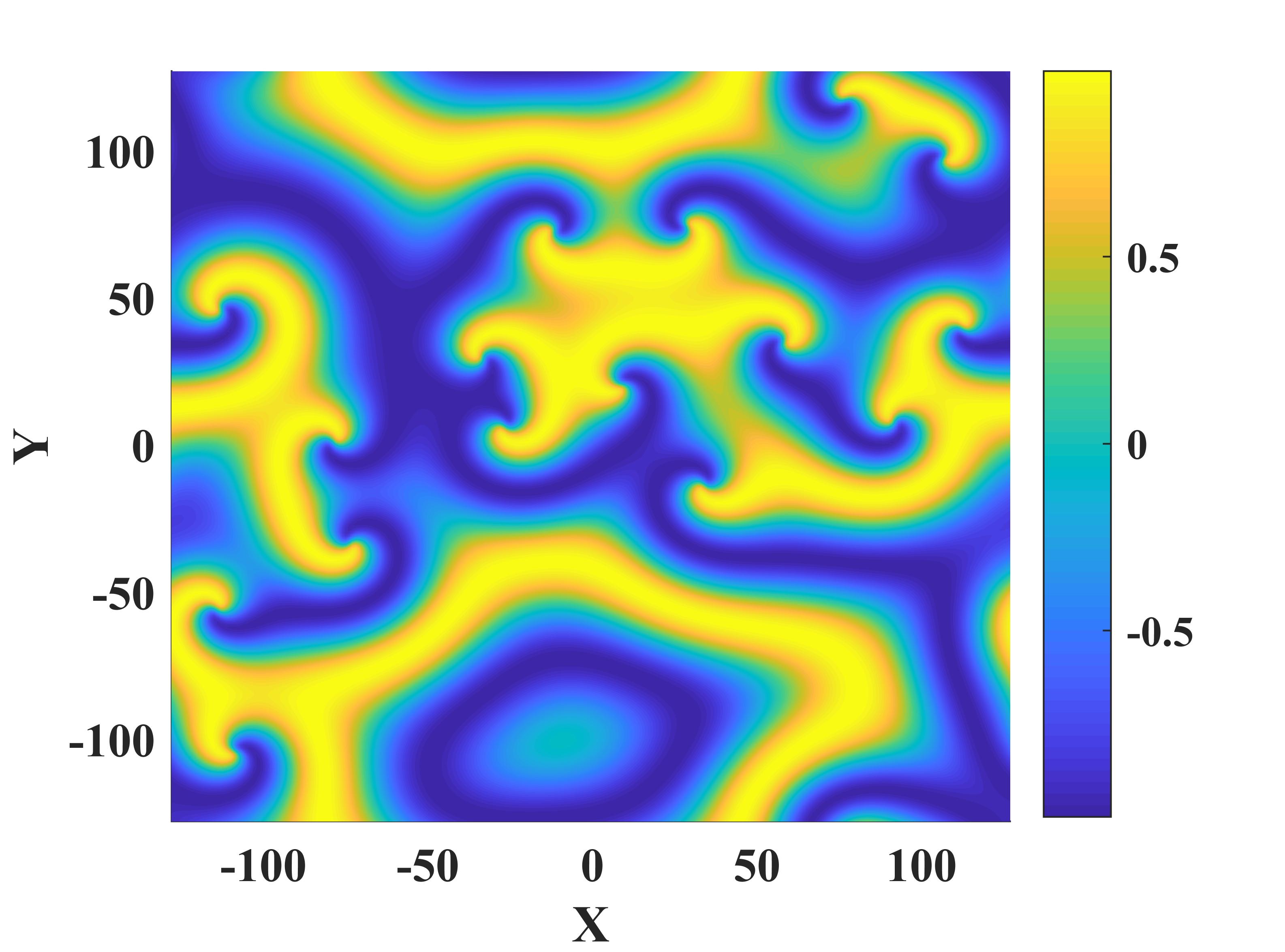}}
		\subfloat[] {\includegraphics[width=0.49\columnwidth]{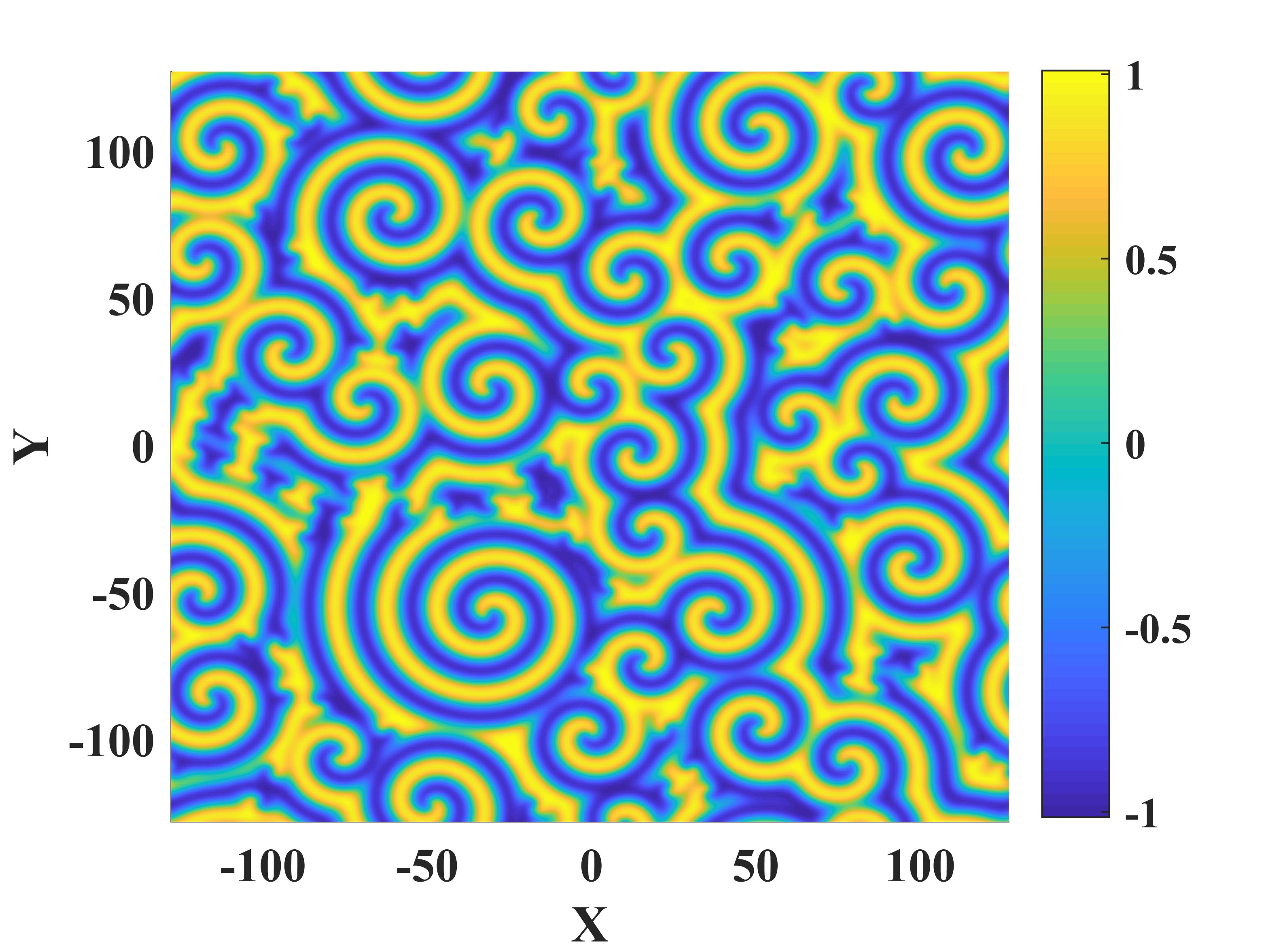}}\\
		\subfloat[] {\includegraphics[width=0.49\columnwidth]{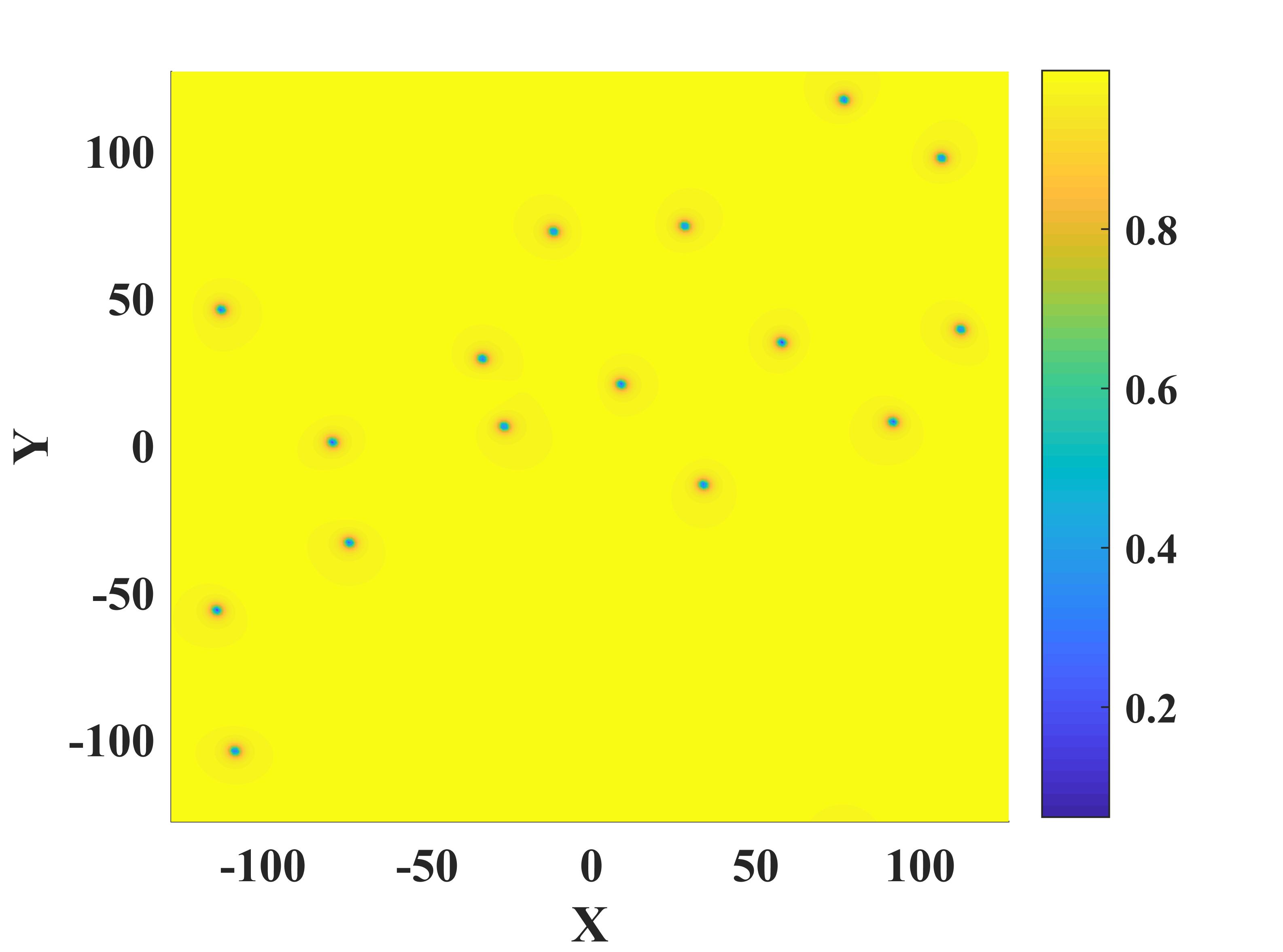}}
		\subfloat[] {\includegraphics[width=0.49\columnwidth]{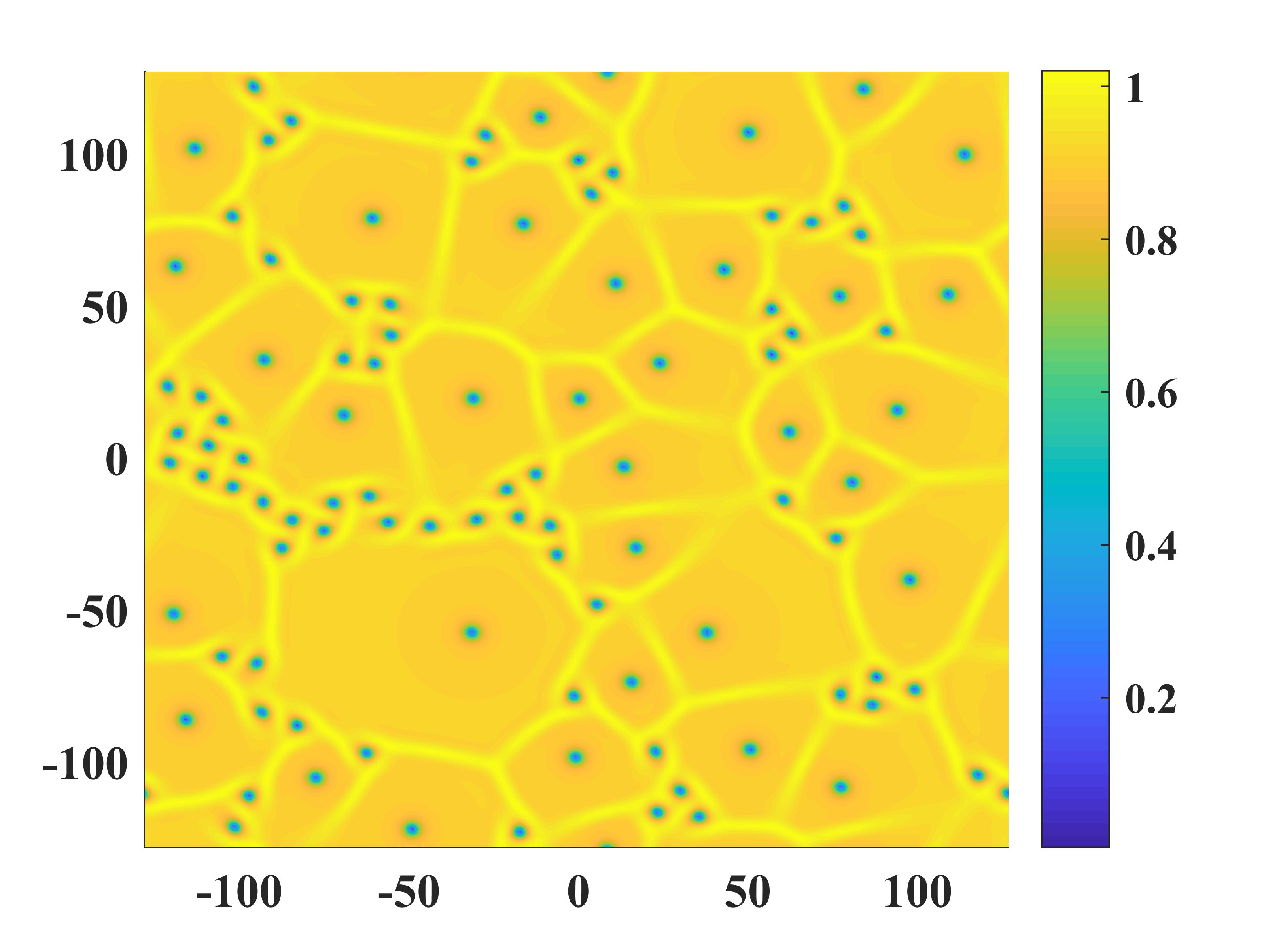}}
	\end{center}	
        \caption{Snapshots the complex Ginzburg-Landau equation~\eqref{CGLERS} for two sets of coefficients with the same $c_2$ but different $c_1$. a) and c) show $Re(W)$ and $|W|$, respectively, for $(c_1, c_2)=(0.2,0.82)$ in the monotonic range. b) and d) show $Re(W)$ and $|W|$, respectively, for $(c_1,c_2)=(-0.33,0.82)$ in oscillatory range. Changing $c_1$ by varying the characteristic time of the short-term memory ($\tau_V$) enables us to cross the oscillatory range line in parameter space \cite{Aranson,AransonK} and help us to change the properties of the spiral waves.}\label{CGLE2}
\end{figure}

The change in the selected spatio-temporal pattern, due to changing the memory property of the system, is not limited to spiral patterns. One can change the coefficient $c_1$ in complex Ginzburg-Landau equation to make a transition, for instance, from defect turbulence regime to phase turbulence regime as well. Also, by focusing on the solution of complex Ginzburg-Landau equation in 1D or 3D, we can study the effect of memory on plane waves or vortex filaments, respectively.  

\section{Summary and discussion} 

Realistic systems possess a finite correlation or scattering time and are generally non-Markovian. In spite of this fact, they are usually treated in the context of Markov approximation where memory is neglected. We studied the effect of memory on the instability and spatio-temporal pattern formation in reaction-diffusion systems and showed that considering memory near the instability might be inevitable. For this purpose, we constructed a general non-Markovian reaction-diffusion system by starting from the generalized master equation under the assumption \eqref{Simp}, to take into account memory effect. 

We considered the Brusselator model as a typical reaction-diffusion system in the presence of memory in the diffusion of the inhibitor. In the limit of short-term memory and for small diffusion coefficients of the inhibitor, the non-Markovian Brusselator model reduced to a set of equations that could be solved analytically. Linear stability analysis indicated that, in the presence of memory, the characteristic equation depends on the characteristic memory kernel decay time. Then, we used the reductive perturbation method near the Hopf instability to investigate the effect of memory on the formation of spatio-temporal patterns. We found that a memory dependent complex Ginzburg-Landau equation governs the amplitude of the critical mode. Since the derived complex-Ginzburg Landau equation was memory dependent, we can change the stability of its solutions by changing the memory property of the system and, thus, control the selected spatio-temporal pattern. These analytical findings were confirmed by numerical simulations of both the non-Markovian Brusselator model and the complex Ginzburg-Landau equation. The results indicate that going beyond the Markovian approximation might be necessary to study the formation of spatio-temporal patterns in many cases and also opened up a new window to the control of these patterns using memory.

We would like to point out that the analytical approach we established here is general. In particular, it is not limited to the non-Markovian Brusselator model. In fact, it can be used to study other reaction-diffusion systems beyond the Markovian approximation in any dimensions. While we showed that considering short-term memory in the diffusion of inhibitor can result in a significant impact on the selected spatio-temporal patterns, considering memory in the diffusion of the activator can lead to an even richer dynamical behavior. In the latter case, the right hand side of equation \eqref{k2} changes and would not necessarily be negative. Therefore, for some parameters, the system can admit nonzero solutions for critical wave numbers ($k_c\neq 0$) where considering them in the system demands more investigations. In addition, one can not readily use reductive perturbation method to study the system in the presence of any nonzero modes. It is because reductive perturbation method is based on the coarse graining of the system to consider important long wavelength modes (with $k=0$ and its neighboring modes). Similarly, while we largely focused on the case of a Hopf bifurcation, the presented framework for studying reaction-diffusion systems in the limit of short-term memories directly gives the conditions for a Turing bifurcation (See equation \eqref{ChMemory} for the case of Brusselator model). This allows one to investigate the effect of short-term memories on Turing patterns using weakly nonlinear analysis \cite{Zhang,Pena}. The mathematical treatment of this case is challenging because of the complicated equations beyond the Markov approximation. Since the main goal of this paper is to provide a proof of concept of how memory can affect the formation of spatio-temporal patterns in analytically tractable way, the detailed investigation of memory in the diffusion of the activator as well the effect of memory on Turing patterns remains an interesting topic for the future.\\ 

JD was financially supported by the Natural Sciences and Engineering Research Council of Canada (NSERC). RT thanks the University of Calgary for their hospitality.

\end{document}